\documentclass[aps,prl,floatfix,twocolumn,footinbib]{revtex4-1}
\usepackage{graphicx,bm,amsmath,amssymb,natbib,url,epsfig,color,hyperref}

\usepackage{enumerate}
\usepackage{amsthm}
\usepackage{mathrsfs}
\usepackage{float}

\DeclareMathOperator{\csch}{csch}

\newcommand{\beginsupplement}{
	\setcounter{table}{0}
	\renewcommand{\thetable}{S\arabic{table}}%
	\setcounter{figure}{0}
	\renewcommand{\thefigure}{S\arabic{figure}}%
	\setcounter{equation}{0}
	\renewcommand{\theequation}{S\arabic{equation}}%
	\setcounter{enumiv}{0}
	\renewcommand{\theenumiv}{S\arabic{enumiv}}%
}

\begin{document}
\title{Fractional Mutual Statistics on Integer Quantum Hall Edges}

\author{June-Young M. Lee}
\affiliation{Department of Physics, Korea Advanced Institute of Science and Technology, Daejeon 34141, Korea}

\author{Cheolhee Han}
\affiliation{Department of Physics, Korea Advanced Institute of Science and Technology, Daejeon 34141, Korea}

\author{H.-S. Sim}\email[]{hssim@kaist.ac.kr}
\affiliation{Department of Physics, Korea Advanced Institute of Science and Technology, Daejeon 34141, Korea}

\date{September 14, 2020}

\begin{abstract}  
{Fractional charge and statistics are hallmarks of low-dimensional interacting systems such as fractional quantum Hall (QH) systems. Integer QH systems are regarded noninteracting, yet they can have fractional charge excitations when they couple to another interacting system or time-dependent voltages. Here, we notice Abelian fractional mutual statistics between such a fractional excitation and an electron, and propose a setup for detection of the statistics, in which a fractional excitation is generated at a source and injected to a Mach-Zehnder interferometer (MZI) in the integer QH regime. In a parameter regime, the dominant interference process involves braiding, via double exchange, between an electron excited at an MZI beam splitter and the fractional excitation. The braiding results in the interference phase shift by the phase angle of the mutual statistics. This proposal for directly observing the fractional mutual statistics is within experimental reach.}      
\end{abstract}
\maketitle



Fractional charge excitations emerge in various interacting systems, including fractional quantum Hall (QH) systems~\cite{fqh0,fqh1} and Luttinger liquids~\cite{tl1,tl-hur2,Kim09,tl2}. The fractional charges have been observed~\cite{fqh2,fqh3,fqh4,fqh5,tl3,tl4}. 

The excitations obey fractional statistics and are called anyons~\cite{Leinaas77,Arovas84,Stern08}.
Upon winding of an anyon around another or their double exchange, their state gains a fractional phase angle in cases of Abelian anyons or evolves into another state (sometimes orthogonal to the initial state) in non-Abelian anyons. 
In fractional QH cases, it has been proposed~\cite{intdevice1} that the statistics can be identified in an interferometer where an anyon propagating along edge channels winds around localized anyons in the QH bulk and the number of the bulk anyons is controlled. 
This strategy is not applicable to detecting the fractional statistics in one-dimensional systems, such as Luttinger liquids~\cite{tl1}, having no bulk anyon to braid of.

\begin{figure}[t]
	\centering
	\includegraphics[width = 0.45 \textwidth]{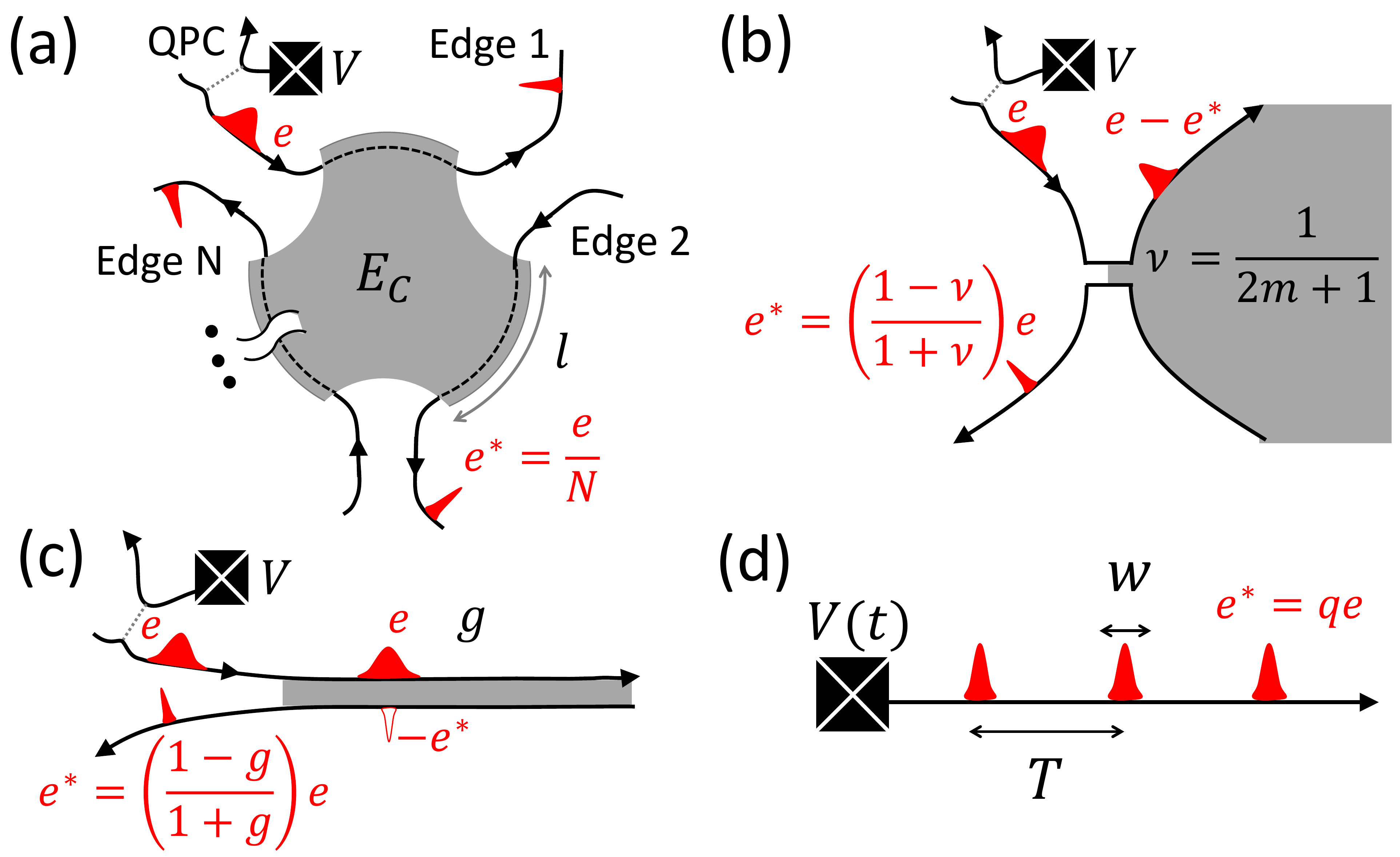}
	\caption{Sources of fractional charges on interger QH edges (solid lines). 
	(a-c) An electron wave packet (thick red peaks) carrying charge $e$  is generated on an edge at a quantum point contact (QPC) by electron tunneling (dotted line) from another edge biased by static voltage $V$.  The electron is fractionalized into charges $e^*$ or $e-e^*$ (thinner peaks) while scattered by an interacting region (shade) of (a) a metallic island coupled to $N$ integer QH edges, (b) fractional QH filling factor $\nu$, or (c) interedge interaction $g$. (d) Fractional charges of period $T$ and spatial width $w$, generated by an Ohmic contact (cross) biased by periodic voltage pulses. } \label{FS}
\end{figure}

Interestingly, fractional charge excitations can be also generated in noninteracting integer QH edge channels, by coupling the channels to interacting systems such as a metallic island (as shown in this work), a fractional QH system~\cite{iqf6}, an interedge-interaction region~\cite{iqf1, iqf2, iqf3, iqf4, iqf5},
or to voltage pulse~\cite{iqf0,aMZI} (see Fig.~\ref{FS}).
They are expected to propagate along the integer QH edge without decay. A question is whether the fractional charges behave as anyons, although they are not of topological order.


In this work, we notice that the fractional charges generated from the sources in Fig.~\ref{FS} obey fractional statistics on integer QH edges, and propose how to detect fractional mutual statistics between a fractional charge and an electron, using a Mach-Zehnder interferometer (MZI)~\cite{Ji03,Roulleau08} in the integer QH regime, which has length difference $\Delta L$ between the two MZI arms. We consider the regime $w \ll \Delta L \lesssim L_\beta$ of the spatial width $w$ of the fractional charge, thermal length $L_\beta = \hbar v \beta/\pi$, thermal energy $1/\beta$, and electron velocity $v$ on the arms. 
The dominant process of this regime consists of (i) injection of a fractional charge from a source of Fig.~\ref{FS} to the MZI and (ii) excitation of an electron at one (the other) MZI beam splitter in one (the other) subprocess.
The interference of the two subprocesses involves braiding, via double exchange, between the fractional charge and the electron.
The resulting phase shift of the interference is determined by the mutual statistics, and identified by comparing the interference with a reference signal from the same setup.




\textit{Fractional mutual statistics.} Excitations of fractional charge $e^* = qe$ 
on an integer QH edge (labeled by $\alpha$) behave as ``particles'' as they form wave packets moving along the chiral edge without deformation~\cite{iqf1, iqf2, iqf3, iqf4, iqf5,iqf0}, although they are not energy eigenstates.
To see their statistics, we identify their creation operator $\eta^\dagger_{\alpha,q} (x)$ on position $x$
in the bosonization~\cite{bos}, where the Hamiltonian $\mathcal{H}_{\text{edge},\alpha} = \frac{\hbar v}{4\pi} \int dx [\partial_x \phi_\alpha(x)]^2$ and electron operator $\psi_{\alpha}(x) \propto e^{i\phi_\alpha(x)}$ of the edge  are represented by a bosonic field  $\phi_\alpha$. Since the excitation carries the charge $qe$, $\eta^\dagger_\alpha (x)$ satisfies
 $[\rho_\alpha (x), \eta_\alpha ^\dagger(x')] = q \delta (x-x') \eta_\alpha ^\dagger (x')$, where $\rho_\alpha (x) = \partial_x \phi_\alpha / (2\pi)$ is the electron density operator.
 Combining it with  the Kac-Moody algebra $[{\phi_\alpha (x)},{ \phi_\beta(x')}] = \pi i \delta_{\alpha\beta}\text{sgn}(x-x')$, we identify $\eta_{\alpha,q}(x) \propto e^{iq \phi_\alpha(x)}$.
 We will show that the fractional charges generated from the sources in Fig.~\ref{FS} are indeed described by $\eta_{\alpha,q}$.

The identification, combined with the Kac-Moody algebra, allows us to find that
position exchange of fractional charges $qe$ and $q'e$ obeys the fractional statistics 
\begin{equation}\label{exchange}
\eta_{\alpha,q}(x)\eta_{\alpha,q'}(x')= 	\eta_{\alpha,q'}(x')	\eta_{\alpha,q}(x)  e^{-i \pi qq'\text{sgn}(x-x')}
\end{equation}
with statistical angle $\pi qq'$. For example, a fractional charge $qe$ and an electron ($q'=1$) satisfy the mutual statistics
$\eta_{\alpha,q}(x)\psi_{\alpha}(x')= 	\psi_{\alpha}(x')	\eta_{\alpha,q}(x)  e^{-i \pi q\text{sgn}(x-x')}$.
with statistical angle $\pi q$.

The angle $\pi q$ is either quantized [Fig.~\ref{FS}(a,b)] or continuously tuned [Fig.~\ref{FS}(c,d)], determined by geometry, interaction strength, or voltage pulse.
This is in stark contrast to the fractional QH cases that the mutual statistics between a fractional particle and an electron is trivial with statistics angle $\pi$~\cite{Stern08}. 
These may be due to the fact that the fractional excitations $\eta_{\alpha,q}$ are not energy eigenstates.

\textit{Fractional charge generation.} We below show how to generate fractional charges using a metallic island in Fig.~\ref{FS}(a) and that they are described by $\eta_{\alpha,q}$. The island is useful~\cite{b0, b1,b2, t1,Landau18,Nguyen20} for simulating Luttinger liquids and multichannel Kondo effects. 

The island couples with $N$ interger QH edges.
The coupling region in each edge has the same length $l$ for simplicity.
The island has the interaction $E_C (n_\text{tot} - n_g )^2$ of excess charge $e(n_\text{tot} - n_g )$.
$E_C$ is the charging energy, $e n_\text{tot}$ is the total charge in the $N$ coupling regions, and $e n_g$ is tuned by gate voltages.
Combining the boson modes $\phi_{\alpha = 1, 2, \cdots, N}$ of the edges, we introduce the charge mode $\tilde{\phi}_c(x) = \sum_{\alpha=1}^N \phi_\alpha(x)/\sqrt{N}$ and neutral modes $\tilde{\phi}_{n, j=1, \cdots, N-1}$ orthonormal to each other.
Then the Hamiltonian  $\mathcal{H}_\text{island} = \mathcal{H}_c + \mathcal{H}_n$ is decoupled into 
the charge part $\mathcal{H}_c= \frac{\hbar v}{4\pi} \int_x (\partial_x \tilde{\phi}_c)^2 + E_C (n_\text{tot} - n_g )^2$
and neutral part $\mathcal{H}_n = \frac{\hbar v}{4\pi} \sum_{j = 1}^{N-1} \int_x (\partial_x \tilde{\phi}_{n,j})^2$.
While the neutral part is noninteracting, the charge mode feels the charging energy.
Since  $n_\text{tot} = \sqrt{N} [\tilde{\phi}_c(l) - \tilde{\phi}_c(0)]/(2\pi)$, where $x = 0$ ($x = l$) is the starting (ending) coordinate of the coupling regions, the ratio of the charging energy to kinetic energy of the charge mode is $\sim N E_C$; the charging energy effectively increases with $N$.


For $N E_C \gg \hbar v/l$, we find~\cite{Supple} charge teleportation with fractionalization. 
In Fig.~\ref{FS}(a), an electron is injected from an edge (different from the edges $\alpha = 1,2,\cdots N$) biased by voltage $V$ to the edge $\alpha = 1$ via tunneling through the QPC. When the electron enters the coupling region ($x=0$), a fractional charge $qe = e/N$ with width $w = \hbar v / (eV)$ {\em immediately} appears at the end ($x=l$) of the coupling region of each edge, independently of $l$. 
\begin{equation}
\begin{split}
\psi_{1}^\dagger(x=0,t) \rightarrow & \prod_{\alpha=1}^N \eta^\dagger_{\alpha,\frac{1}{N}} (x = l,t).
\end{split}
\label{teleport}
\end{equation}
The teleportation of the electron into the $N$ fractional charges happens due to the edge chirality and the large charging energy $N E_C$ that prohibits charge modulation inside the coupling regions. It can be seen~\cite{Supple} from the equation of motion of the charge mode, 
 $\tilde{\phi}_c(x ,t)=  \tilde{\phi}_c^{(0)}(x-l,t) + 2\pi n_g / \sqrt{N}$ for $x>l$.
$\tilde{\phi}_c^{(0)}$ denotes the charge mode of the $E_C = 0$ case. 
The teleportation is accompanied [not shown in Eq.~\eqref{teleport}] by 
neutral excitations moving inside the coupling regions. 
Since the neutral excitations decay out before moving out of the coupling regions~\cite{b3}, the setup of Fig.~\ref{FS}(a) generates fractional charges $\eta^\dagger_{\alpha, 1/N}$, obeying Eq.~\eqref{exchange}, on each edge.
The teleportation of the $N=1$ case (without the fractionalization) has been proposed~\cite{t2} and experimentally supported~\cite{t1}.

 
 
Similarly, the setups in Fig.~\ref{FS}(b-c) generate fractional charges, described~\cite{Supple} by $\eta_{\alpha,q}$ in a stochastic way utilizing a QPC and an interaction region.
The fraction $q$ is governed by the filling factor or interedge interaction.

By contrast, the setup in Fig.~\ref{FS}(d) generates fractional charges on demand~\cite{iqf0}.
Here a voltage pulse (per period) $V(t)$ of Lorentzian shape with temporal width $w/v$ satisfying $e \int V(t) dt = 2  \pi q \hbar$
generates a charge $qe$ with  width $w$. 
When $q < 1$, it generates a fractional charge. Although the fractional charge is composed of many electron-hole pair excitations~\cite{iqf0}, it is described by $\eta_{\alpha,q}$ and satisfies Eq.~\eqref{exchange}. 
This can be seen from the time evolution operator $\mathcal{T}e^{-\frac{i}{\hbar}\int_{t'} \mathcal{H}_V(t')}$ in the interaction picture under the Hamiltonian $\mathcal{H}_V(t) =  e V(t)\int_{-\infty}^0 \rho_\alpha (x'-vt) dx'$ due to the voltage pulse applied at, saying, $x<0$ on edge $\alpha$, where $\mathcal{T}$ is the time ordering.
Since the voltage pulse is well localized (like a delta function), the evolution operator reduces to $\eta^\dagger_{\alpha, q} (x)$, $\mathcal{T}e^{-\frac{i}{\hbar}\int_{t'} \mathcal{H}_V(t')}  \to \eta^\dagger_{\alpha, q} (x)$, on length scales $\gg w$.
This generation is beneficial as one can tune $q$ and the number of charges (per period).


\begin{figure*}[t!]
	\centering
	\includegraphics[width = 0.9\textwidth]{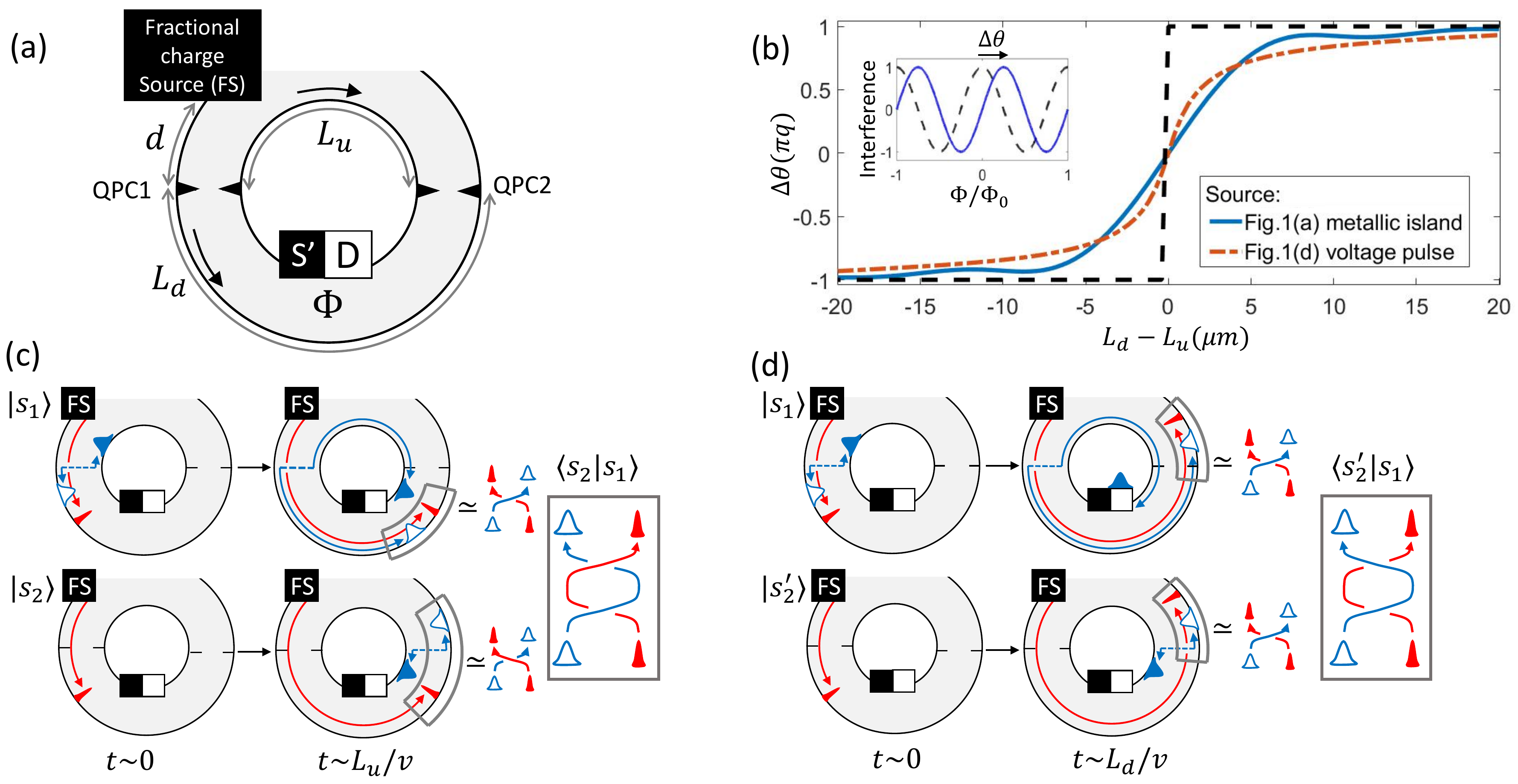}
	\caption{MZI for detecting the fractional mutual statistics. 
	(a) The MZI  has two (upper and lower) integer QH edge channels (the boundaries of the shaded region, with chirality indicated by thick arrows) and two QPCs (its two beam splitters).  The length $L_{u}$ ($L_d$) of the upper (lower) MZI arm is chosen as $L_d > L_u$. The source (one of Fig.~\ref{FS}) generates fractional charges $qe$ on the lower edge, so the charges are injected to the MZI. Dilute injection leading to only one or no fractional charge inside the arms at any time is considered. Interference signals are detected at reservoir D.  
	(b) Phase shift $\Delta \theta$ of the Aharonov-Bohm interference signal versus $L_d - L_u$.
	 It approaches to the statistics angle $\pm \pi q$ at large $\Delta L = |L_d-L_u|$, and to the black dashed line as $w \rightarrow 0$. 
	 The signals are interference conductance (blue solid) computed with a metallic island of $N=3$ ($q=1/3$), $V = 40 \, \mu$V ($w = 1.6\mu m$), $E_C \to \infty$~\cite{realistic}, and inteference current (orange dash-dot) with voltage pulses of $q = 1/3$, temporal width $w/v = 3$ ps, period $T= 250$ ps.
	Temperature 20 mK and $v = 10^5$ m/s are chosen.
	Inset: The phase shift is obtained by comparing the signal (solid) with a reference (dashed). The reference interference is obtained with turning off the fractional charge source and applying a small voltage to reservoir S$'$.
	(c-d) Interference processes. In a subprocess $|s_1 \rangle$, an electron (wide blue packet) and a hole (empty) are pair-excited (dashed line) at QPC1 at time $t = 0$, after a fractional charge (thin red) passes QPC1. In $|s_2 \rangle$ (resp. $|s_2' \rangle$), a pair excitation happens at QPC2 at $t \sim L_u/v$ (resp. $t \sim L_d/v$) before (resp. after) the fractional charge passes QPC2. The interference $\langle s_2 | s_1 \rangle$ between $|s_1 \rangle$ and $|s_2\rangle$ involves double exchange between the electron and fractional charge, while the interference $\langle s_2' | s_1 \rangle$  does not (see the boxes).
	 }\label{MZI}
\end{figure*}  

\textit{Detecting the mutual statistics. } 
The statistics in Eq.~\eqref{exchange} can be detected by injecting fractional charges to an MZI.
In Fig.~\ref{MZI} we illustrate the case of dilute injection, which is achieved by tuning  the QPC or voltage pulse in Fig.~\ref{FS}. 
The MZI is in the regime of $w \ll \Delta L \lesssim L_\beta$, and its two QPCs are in the electron-tunneling regime.
The conventional interference of the MZI occurs~\cite{Supple} as interference of processes with and without splitting of an injected fractional charge $qe$ at QPC1 into $e$ in the upper arm and $-(1-q)e$ in the lower arm.
This interference is however negligible since the width $w$ of the fractional charge is much shorter than the arm length difference $\Delta L$.
Instead there is a new interference process in which an electron braids, via double exchange, with the fractional charge, with the help of electron tunneling at the QPCs; the splitting of the fractional charge does not happen in the new process.
This interference is visible when the thermal length is not too short ($\Delta L \lesssim L_\beta$).

To illustrate the new process, we first discuss the corresponding process in thermal equilibrium with no fractional charge injection. We consider an electron on the lower edge of the MZI, located at QPC1 (whose location is $x=0$ on both edges) at time $t=0$. In a subprocess $|s_1 \rangle_0$, this electron jumps to the upper edge at QPC1 via tunneling at $t=0$.
The resulting state is $|s_1 \rangle_0 = \psi_u^\dagger(0,0)|\rangle_u  \psi_d(0,0)|\rangle_d$, where $\psi_{u(d)}^\dagger(x,t)$ creates an electron on the Fermi sea $|\rangle_{u (d)}$ of the upper  (lower) edge.
The electron arrives at QPC2 at $t = L_u/v$.
In another subprocess $|s_2\rangle_0$ (resp. $|s_2'\rangle_0$), which interferes with $|s_1\rangle_0$ with the largest overlap, the electron moves along the lower edge 
and electron tunneling from the lower to upper edge happens at QPC2 at $t = L_u/v$ (resp. $t = L_d/v$); $|s_2\rangle_0=\psi_u^\dagger(L_u,L_u/v)|\rangle_u  \psi_d(L_d,L_u/v)|\rangle_d$ and $|s_2'\rangle_0=\psi_u^\dagger(L_u,L_d/v)|\rangle_u  \psi_d(L_d,L_d/v)|\rangle_d$.
The interference between $|s_1 \rangle_0$ and $|s_2 \rangle_0$ cancels 
that between $|s_1 \rangle_0$ and $|s_2' \rangle_0$, $\langle s_2' | s_1 \rangle_0 + \langle s_2 | s_1 \rangle_0 = 0$. This is  proved~\cite{Supple} with the Fermi statistics and the chirality $\psi(x,t) = \psi(x-vt)$. The full cancellation is natural in equilibrium.

 
 
 
The full cancellation does not happen when a fractional charge $qe$ is injected to the MZI.  The fractional charge is generated at $x = -d$ on the lower edge at $t=-d/v +t_0$, much earlier than the other events ($d$ is assumed large).
Then the subprocesses are modified,
\begin{eqnarray}
|s_1 \rangle &=& \psi_u^\dagger(0,0|\rangle_u  \psi_d(0,0)\eta_{d,q}^\dagger(-d,-\frac{d}{v} +t_0)|\rangle_d, \label{qinj} \\
|s_2 \rangle &=& \psi_u^\dagger(L_u,\frac{L_u}{v})|\rangle_u  \psi_d(L_d,\frac{L_u}{v})\eta_{d,q}^\dagger(-d,-\frac{d}{v} +t_0)|\rangle_d, \nonumber \\ 
|s_2' \rangle &=& \psi_u^\dagger(L_u,\frac{L_d}{v})|\rangle_u  \psi_d(L_d,\frac{L_d}{v})\eta_{d,q}^\dagger(-d,-\frac{d}{v} +t_0)|\rangle_d, \nonumber
\end{eqnarray}
such that electron tunneling happens at a QPC as in the previous case,
while the fractional charge stays on the lower edge, passing
QPC1 at time  $t_0 \neq 0, (L_u - L_d) / v$ so that it does not overlap with the electron.  
In the domain of $t_0$ between 0 and  $(L_u - L_d) / v$, the fractional charge passes QPC1 {\em before} the electron tunneling in $|s_{1 }\rangle$ and passes QPC2 {\em after} the electron tunneling in $|s_{2}\rangle$ in the setup with $L_d > L_u$ [Fig.~\ref{MZI}(c)].
Hence, an exchange between the fractional charge and electron happens in $|s_1\rangle$ 
in the direction opposite to $|s_2 \rangle$.
The interference between $|s_1 \rangle$ and $|s_2\rangle$ gains, in comparison with the case of no fractional charge injection, a phase factor $e^{-i2\pi q}$ of the mutual statistics angle due to double exchange [see Eq.~\eqref{exchange} with $q'=1$]. 
In the same $t_0$ domain of the $L_d < L_u$ case, the events happen in reverse time ordering, and the interference gains $e^{i2\pi q}$.
This is shown as
\begin{equation} \label{PhShift}
\langle s_{2 } | s_{1 } \rangle = \frac{1}{2 \pi a}  \langle s_{2 } | s_{1 } \rangle_0 e^{\mp i2\pi q}
\end{equation}
with  $e^{- i2\pi q}$ for $L_d > L_u$, $e^{i2\pi q}$ for $L_d < L_u$, and 
$2\pi a$ from the short-length cutoff of $\eta_{d,q}$.
In the other domain of $t_0$ [but $t_0 \ne 0, (L_u - L_d)/v$], the double exchange does not happen, $\langle s_{2 } | s_{1 } \rangle = \langle s_{2 } | s_{1 } \rangle_0 / (2 \pi a)$.
By contrast, $|s_2' \rangle$ has an exchange in the same direction with $|s_1 \rangle$ for any $t_0$ [Fig.~\ref{MZI}(d)], hence, $\langle s_2' | s_{1 } \rangle = \langle s_2' | s_{1 } \rangle_0 / (2 \pi a) = - \langle s_2 | s_1 \rangle_0 / (2\pi a)$.
Thus, the two interferences cancel only partially, $\langle s_2 | s_1 \rangle + \langle s_2' |s_1 \rangle \propto  (e^{\mp i 2 \pi q} -1)  \langle s_{2 } | s_{1 } \rangle_0$ in the domain of $t_0$ between 0 and  $(L_u - L_d) / v$.

  


The partial cancellation due to the mutual statistics has direct consequence in interference signals
[differential conductance for the sources in Fig.~\ref{FS}(a-c) and current for Fig.~\ref{FS}(d)] measured at detector D.
For $w \ll \Delta L$,
the processes and their complex conjugate lead~\cite{Supple} to
\begin{eqnarray}
\text{Interference signal} & \propto & \int dt_0 \, \text{Re}[\langle s_{2 } | s_{1} \rangle +\langle s_{2 }' |  s_{1 } \rangle]	\nonumber	\\
& \propto & \frac{\Delta L}{\beta v}\Lambda_\beta \, \text{Re}[\pm i(e^{\mp  i2\pi q} -1)e^{-2\pi i\frac{\Phi}{\Phi_0}}] \nonumber \\
& \propto & \frac{\Delta L}{\beta v}\Lambda_\beta   \sin (\pi q) \cos(2\pi\frac{\Phi}{\Phi_0} \pm \pi q),   \label{main}
\end{eqnarray} 
with the magnetic flux $\Phi$ enclosed by the MZI, the flux quantum $\Phi_0 = h/e$,  the thermal dephasing factor $\Lambda_\beta = \csch(\Delta L/L_\beta)$, and the sign $\pm$ determined by $\text{sgn}(L_d-L_u)$. 
The second line is obtained, using Eq.~\eqref{PhShift} and $\langle s_2 | s_1 \rangle_0 \propto \pm i (\Lambda_\beta / \beta) e^{-2\pi i \frac{\Phi}{\Phi_0}}$.
The factor $\Delta L / v $ comes from the $t_0$ domain of the partial cancellation, showing that the double exchange is more probable at larger $\Delta L / w$.
The factor $\sin(\pi q)$ indicates full cancellation between $\langle s_2 | s_1\rangle$ and $\langle s_2' | s_1 \rangle$ when an electron, instead of a fractional charge, is injected ($q=1$).
Due to the thermal dephasing, the signal is visible at $\Delta L \lesssim L_\beta$.
Equation~\eqref{main} shows that the interference pattern is shifted by $\pi q$ due to the mutual statistics,
which can be read out by comparing the pattern with a reference [Fig.~\ref{MZI}(b)]. 

The interference with the double exchange is compared with the conventional process of the MZI. The conventional interference occurs with $t_0 = 0$ or $(L_u - L_d)/v$ such that electron tunneling happens at QPC1 or QPC2 (leaving a fractional hole behind)  when a fractional charge arrives at the QPC~\cite{Supple}. This process is negligible when $\Delta L / w \gg 1$.
This is confirmed in the calculation [Fig.~\ref{MZI}(b)] including all the processes and obtained with experimentally feasible parameters and the Keldysh Green functions.
The calculation becomes identical to the result of Eq.~\eqref{main} at $\Delta L \gg w$. 

When one changes the voltage pulse shape (varying the pulse period $T$ or applying a group of pulses) in Fig.~\ref{FS}(d), one can further control the number of injected fractional charges inside the MZI. 
Then in the interference $\langle s_2 | s_1 \rangle$, the electron can braid more than one fractional charge, $n = \lfloor \frac{\Delta L}{vT} \rfloor$ or $n+1$ fractional charges with probability $p_n=  \lceil \frac{\Delta L}{vT} \rceil-\frac{\Delta L}{vT}$ or $p_{n+1}=1-p_n$. Here, $\lfloor x \rfloor $ is the largest integer $\le x$ 
and $\lceil x \rceil $ is the smallest integer $\ge x$.
Another interference $\langle s_2' | s_1 \rangle$ has no braiding. For $w \ll \Delta L \lesssim L_\beta$, the time-averaged interference current is found~\cite{Supple},
\begin{equation}\label{resVT}
\overline{I_\text{D}^\text{int}} \propto |f| \Lambda_\beta \cos(2\pi  \frac{ \Phi}{\Phi_0} -\arg f),	 
\end{equation}
where $f = \pm i (p_ne^{\mp i2\pi qn }	+ p_{n+1}e^{\mp i2\pi q (n+1)}-1 )$ corresponds to the factor $\pm i \Delta L (e^{\mp i 2 \pi q} -1)$ in Eq.~\eqref{main}. Equation~\eqref{resVT} reduces to Eq.~\eqref{main} for $vT > \Delta L$.


\textit{Concluding remark.}
We demonstrated that fractional charges on integer QH edges obey the fractional mutual statistics. 
Our proposal for directly detecting the statistics is within experimental reach: The setups in Fig.~\ref{FS} are experimentally available. The MZI with long coherence length has been realized many times. 
The parameters used in Fig.~\ref{MZI} are realistic; in Fig.~\ref{MZI} the phase shift $\pm \pi q$ is obtained for $\Delta L \sim 10 \mu\textrm{m}$, with which the interference visibility is expected larger than 0.7. 

We note that environment effects can cause the same amount of an additional phase shift in both the interference signal and the reference. Hence, they do not affect the detection of the phase shift by the mutual statistics. Moreover, the dynamical phase of injected fractional charges does not cause a phase shift in Eq.~\eqref{main}, since a fractional charge propagates the same distance in the interfering subprocesses.
There is one obstacle to this direction of detecting the mutual statistics, when the MZI is realized with $\nu=2$ QH edges.
There, interedge Coulomb interaction can result in additional unwanted fractionalization~\cite{iqf3,iqf4,iqf5}.
One can avoid this by gapping out the inner edge as done recently~\cite{macrocohe}.


Our main process with the double exchange is nontrivial, existing with the help of the fractional statistics.
The process has been unnoticed before, maybe because the full cancellation between $\langle s_2 | s_1 \rangle$ and $\langle s_2' |s_1 \rangle$ is restored (causing no response in observables) when fractional charges are replaced by electrons ($q = 1$). 
 
The process relies on the double exchange between two particles on QH edges
rather than braiding with anyons in QH bulk. Such double exchange will be useful also for detecting anyonic statistics
in fractional QH interferometries~\cite{tvb1,tvb3} or anyon collision setups~\cite{tvb2,Bartolomei}.

 Our work suggests to explore fractional statistics in noninteracting systems, which will be useful for engineering anyons and for detecting, nonlocally with an MZI, the information of the interaction regions (e.g., the interedge interaction).
It will be valuable to generalize our work to
the fractional statistics of Luttinger liquids.





We thank Fr\'{e}d\'{e}ric Pierre for insightful discussions which motivated us to investigate the present work, and Anne Anthore and Wanki Park for helpful discussions. This work is supported by Korea NRF (SRC Center for Quantum Coherence in Condensed Matter, Grant No. 2016R1A5A1008184). JYML is also supported by Korea NRF (NRF-2019-Global Ph.D. Fellowship Program).


\cleardoublepage
\pagebreak
\beginsupplement

\widetext
\begin{center}
	\textbf{\large Supplemental Materials: Fractional Mutual Statistics on Integer Quantum Hall Edges}
	\linebreak \\
	June-Young M. Lee, Cheolhee Han,  and H.-S. Sim* \\
	
	\textit{\small Department of Physics, Korea Advanced Institute of Science and Technology, Daejeon 34141, Korea}
\end{center}

\setcounter{equation}{0}
\setcounter{figure}{0}
\setcounter{table}{0}
\setcounter{page}{1}
\makeatletter
\renewcommand{\thesection}{S\Roman{section}}
\renewcommand{\theequation}{S\arabic{equation}}
\renewcommand{\thefigure}{S\arabic{figure}}
\renewcommand{\bibnumfmt}[1]{[S#1]}
\renewcommand{\citenumfont}[1]{S#1}

\section{Operator of fractional charges}

We show in details that the fractional charges generated by the sources in Fig.~1(a-c) are described by the operator $\eta_{q,\alpha}$.
We first consider the setup in Fig.~1(a). The Hamiltonian for the charge mode $\tilde{\phi}_c$ is  $\mathcal{H}_c= \frac{\hbar v}{4\pi} \int_x (\partial_x \tilde{\phi}_c)^2 + NE_C (\tilde{\phi}_c(l)-\tilde{\phi}_c(0) - 2\pi n_g/\sqrt{N} )^2/(2\pi)^2$ (see the main text). The solution of the equation of motion of $\phi_c$ for $x>l$ is 
\begin{equation}
\begin{split}
\tilde{\phi}_c(x>l,t)= & \tilde{\phi}^{(0)}_c(x,t) - i\int \frac{dq}{2\pi}  \frac{2g_C  \frac{\sin^2 ql/2}{ql/2}}{1+g_C\frac{\sin ql/2}{ql/2}e^{iql/2}}\tilde{\phi}^{0}_c(q)e^{iq(x-vt)}	+\frac{g_C}{1+ g_C}\frac{2\pi n_g}{\sqrt{N}}	
\end{split}
\end{equation}
where $g_C = \frac{NE_C}{\pi \hbar v/l}$ is dimensionless strength of the effective Coulomb interaction. Hence, for $NE_C  \gg \hbar v/l$, we get
\begin{equation}
\begin{split}
\tilde{\phi}_c(x>l,t)= & \tilde{\phi}_c^{(0)}(x-l,t) + 2\pi \frac{n_g}{\sqrt{N}},			\\
\end{split}
\end{equation}
implying teleportation of the charge mode by the distance $l$. On the other hand, the neutral modes $\tilde{\phi}_{nj}$ propagate freely. 
Combining these, we find that the boson field of the edge $\alpha=1$ at the entrance $x=0$ of the coupling region satisfies $\phi_1(0^-,t) = \sum_{\alpha = 1}^N \frac{1}{N}\phi_\alpha(l,t) + (1-\frac{1}{N})\phi_1(0^+,t) - \sum_{\alpha = 2}^N \frac{1}{N}\phi_\alpha(0^+,t)$. Exponentiating this, we find that the electron creation operator at $x=0$ on the edge $\alpha = 1$ reduces to
\begin{equation}
\psi_{1}^\dagger(x=0,t) \rightarrow  \prod_{\alpha=1}^N \eta^\dagger_{\alpha,\frac{1}{N}} (l) \left(\eta^\dagger_{1,1-\frac{1}{N}}(0,t)\prod_{\alpha=2}^N \eta_{\alpha,\frac{1}{N}}(0,t)	\right),
\end{equation}
where the fractional excitations inside (outside) the parenthesis are described by the neutral modes (the charge mode). Neglecting the neutral modes as they decay out before moving out of the coupling regions, we arrive at Eq.~(2).

We next consider the setup in Fig.~1(c).
There, the left-moving and right-moving edge channels, described by the boson fields $\phi_L$ and $\phi_R$, are decoupled at $x<0$ and coupled at $x>0$ (Luttinger-liquid region) by interedge Coulomb interaction $\mathcal{H}_\text{int} = -\frac{\hbar}{4\pi} 2V \int^\infty_0 dx \partial_x \phi_L(x) \partial_x\phi_R(x)$. Note that the electron operator and fractional-charge operator of the left-moving edge channel
have the same form, $\psi_L(x) \propto e^{i\phi_L(x)}$ and $\eta_{L,q}(x) \propto e^{iq\phi_L(x)}$, with those of the right-moving edge channel, while the Kac-Moody algebra and electron density operator have an additional minus sign, $[{\phi_L (x)},{ \phi_L(x')}] = -\pi i\, \text{sgn}(x-x')$ and $\rho(x) = -\partial_x\phi_L(x)/(2\pi)$. Solving the total Hamiltonian, we find that the region of $x>0$ has
the interaction parameter (called the Luttinger-liquid parameter) $g= \sqrt{(v-V)/(v+V)}$, renormalized velocity $u =  \sqrt{v^2-V^2}$, and right-moving eigen-mode $\tilde{\phi}_R(x)= (\phi_R(x) -r \phi_L(x))/\sqrt{1-r^2}$, and left-moving eigen-mode $\tilde{\phi}_L(x) =( r\phi_R(x) + \phi_L(x) )/\sqrt{1-r^2}$, where $r = (1-g)/(1+g)$. The eigenmodes follow the commutator relation of $[{\tilde{\phi}_{R(L)} (x)},{ \tilde{\phi}_{R(L)}(x')}] = \pm\pi i\, \text{sgn}(x-x')$.

To see the fractionalization of a right-moving electron at $x=0$, we use the continuity of the boson fields $\phi_{L(R)}(0^-,t) = \phi_{L(R)}(0^+,t)$ at $x=0$, which leads to $\phi_R(0^-,t) = r\phi_L(0^-,t)+\sqrt{1-r^2}\tilde{\phi}_R(0^+,t)$.
Therefore, the right-moving electron is fractionalized at $x=0$ into the reflected part described by  $ r\phi_L(0^-,t)$ and the transmitted part by $\sqrt{1-r^2}\tilde{\phi}_R(0^+,t)$.  The fractional charge generation at $x<0$ (the noninteracting region) is thus shown as
\begin{equation}
\psi^\dagger_R(x=0,t) \rightarrow \eta^\dagger_{L,r}(0,t).
\end{equation} 

An electron and a quasiparticle of the Luttinger liquid are expressed as $ e^{i\tilde{\phi}_{L(R)}/\sqrt{g}}$ and $ e^{i\sqrt{g}\tilde{\phi}_{L(R)}}$ respectively, and they exactly correspond to those of the Laughlin state for $g = \nu = \frac{1}{2m+1}$. The mapping implies the fractionalization in the setup in Fig.~1(b) is mathematically identical to that of Fig.~1(c).

\section{Overlap between the interference subprocesses \& Interference signal}

We provide the calculation of the overlap between the subprocesses $|s_1\rangle$, $|s_2 \rangle$, and $|s_2'\rangle$, and the mathematical expression of the interference signal. Hereafter we use convention of $\hbar \equiv 1$.

We use the normalized version of the fractional-charge operator,  $\eta_{\alpha,q} = e^{iq\phi_{\alpha,q}}/\sqrt{2\pi a}$, where $a$ is the short-length cutoff. The two-point correlation function of the operators is written as
\begin{equation}
\langle \eta_{q}^\dagger (x,t) \eta_q (0,0)\rangle= \frac{1}{2\pi a}\frac{1}{[\frac{\beta v}{\pi a}\sin(\frac{\pi}{\beta v}(a-i(x-vt))]^{q^2}},
\end{equation}
and reduces to the electron correlation function for $q=1$. 
This is consistent with the exchange rule of Eq. (1) of the main text, 
as $\langle \eta_{q}^\dagger (x,t) \eta_q (0,0)\rangle = \langle  \eta_q (0,0) \eta_{q}^\dagger (x,t)\rangle e^{i\pi q^2 \text{sgn}(x-vt)}$ for $|x-vt| \gg a$. 

When no fractional charge is injected to the MZI, the interference, corresponding to the main interference process in the presence of the fractional-charge injection,  vanishes $\langle s_2' |s_1 \rangle_0 + \langle s_2 |s_1 \rangle_0 = 0$ when no fractional charge is injected to the MZI.
In this case, the overlap between the subprocesses is written, using the correlation function, as
\begin{equation}
\begin{split}
\langle s_2 |s_1 \rangle_0 & = \langle \psi^\dagger_d(L_d-L_u, 0)\psi_d(0,0) \rangle \langle \psi_u(0,0)\psi_u^\dagger (0,0) \rangle = \frac{1}{4\pi^2 a} \frac{e^{-2\pi i\frac{\Phi}{\Phi_0}}}{\frac{\beta v}{\pi}\sin[\frac{\pi}{\beta v}(a -i(L_d-L_u))]}, \\
\langle s_2' |s_1\rangle_0  & = \langle \psi^\dagger_d(0,0)\psi_d (0,0) \rangle \langle \psi_u(L_u-L_d,0)\psi^\dagger_u(0,0) \rangle= \frac{1}{4\pi^2 a} \frac{e^{-2\pi i\frac{\Phi}{\Phi_0}}}{\frac{\beta v}{\pi}\sin[\frac{\pi}{\beta v}(a +i(L_d-L_u))]}, 
\end{split}
\end{equation}
hence $\langle s_2' |s_1\rangle_0  =  - \langle s_2 |s_1 \rangle_0$ for $a \rightarrow 0$. 

In contrast, in the presence of the fractional-charge injection, the main interference process, involving the double exchange between a fractional charge and an electron, has the overlap between its subprocesses

\begin{equation}
\begin{split}
\langle s_2 |s_1 \rangle & = \langle \eta_{d,q}(-vt_0,0)\psi^\dagger_d(L_d-L_u,0 )\psi_d(0,0)\eta^\dagger_{d,q}(-vt_0,0) \rangle \langle \psi_u(0,0)\psi_u^\dagger (0,0) \rangle = \frac{1}{8\pi^3a^2} \frac{e^{i\pi q [\text{sgn}(t_0)-\text{sgn}(t_0-(L_u-L_d)/v)]}e^{-2\pi i\frac{\Phi}{\Phi_0}}}{\frac{\beta v}{\pi}\sin(\frac{\pi}{\beta v}(a -i(L_d-L_u))}, \\
\langle s_2' |s_1\rangle  & = \langle \eta_{d,q}(-vt_0,0)\psi^\dagger_d(0,0)\psi_d (0,0) \eta_{d,q}^\dagger(-vt_0,0)\rangle \langle \psi_u(L_u-L_d,0)\psi^\dagger_u(0,0) \rangle= \frac{1}{8\pi^3 a^2} \frac{e^{-2\pi i\frac{\Phi}{\Phi_0}}}{\frac{\beta v}{\pi}\sin(\frac{\pi}{\beta v}(a +i(L_d-L_u))}.
\end{split}
\end{equation}

Hence, 

\begin{equation}
\begin{split}
\text{Interference signal} &\propto \int dt_0\text{Re}[\langle s_{2 } | s_{1} \rangle +\langle s_{2 }' |  s_{1 } \rangle]\\ &= \frac{1}{8\pi^3a^2 }\frac{\Delta L}{ v }\text{Re}[\frac{(e^{\mp i2\pi q} -1)e^{-2\pi i\frac{\Phi}{\Phi_0}}}{\frac{\beta v}{\pi}\sin(\frac{\pi}{\beta v}(a -i(L_d-L_u))}	] = \frac{1}{2\pi^2a^2 v}\frac{\Delta L}{\beta v}\Lambda_\beta \sin \pi q\cos(2\pi \frac{\Phi}{\Phi_0} \pm \pi q),
\end{split}
\end{equation}

where $\Lambda_\beta = \csch(\Delta L/L_\beta)$.
This results in Eq. (5) of the main text. 

\section{MZI coupled to a metallic island}

\begin{figure}[h!]
	\centering
	\includegraphics[width = 0.8 \textwidth]{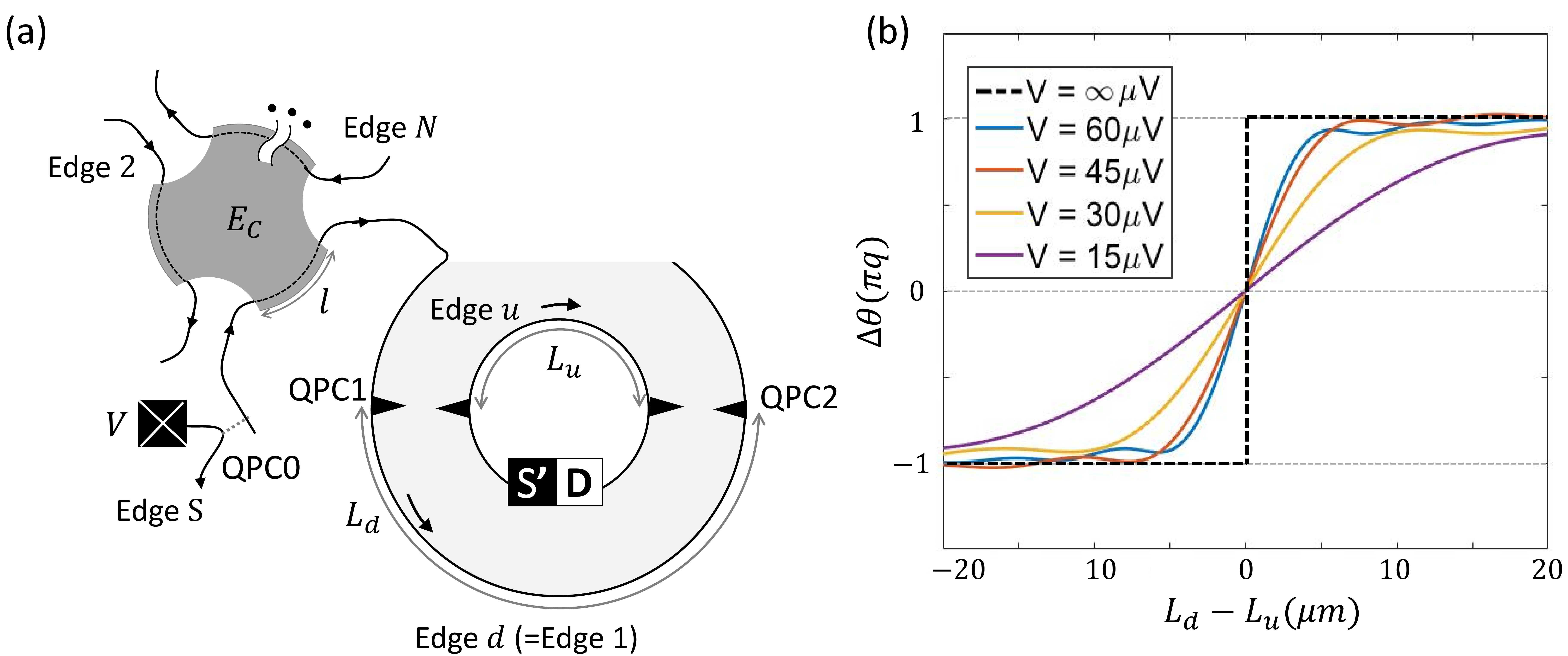}
	\caption{(a) Setup for the MZI combined with a fractional-charge source formed by the metallic island. Bias voltage $V$ is applied on the edge S. (b) The phase shift of the interference conductance for various strengths of the bias voltage $V$, computed with temperature at 20 mK, $v = 10^5$ m/s, and $q= 1/N = 1/3$.}\label{TVBsetup}
\end{figure}

We provide a detailed calculation of the interference conductance in the setup where the asymmetric MZI (shown in Fig. 2 of the main text) is combined with the fractional-charge source formed by a metallic island (shown in Fig. 1). The full setup is shown in Fig.~\ref{TVBsetup} of this supplementary material.

The total Hamiltonain of the setup is written as
\begin{equation}
\begin{split}
\mathcal{H}= &	\mathcal{H}_\text{edge,S} + \mathcal{H}_\text{inj} + \mathcal{H}_\text{island}  + \mathcal{H}_{\text{edge},u}+\mathcal{H}_\text{MZI-QPC}.	 
\end{split}
\end{equation}
Here, $\mathcal{H}_\text{edge,S}$ is the Hamiltonian for the edge S, on which the bias voltage $V$ is applied (see Fig.~\ref{TVBsetup}).
An electron jumps, via tunneling at the QPC0, from the edge S to one (saying edge $\alpha = 1$)  of the edges connected to the metallic island.
This is described by the ``injection'' Hamiltonian $\mathcal{H}_\text{inj} =   \mathcal{T}_0 + \mathcal{T}_0^\dagger$, where $\mathcal{T}_0 = \Gamma_0 \psi_{1}^\dagger(-\frac{d}{2})\psi_\text{S}(0)$ and the tunneling amplitude $\Gamma_0 = \gamma_0 e^{ieV t/\hbar}$ includes the effect of the bias voltage $V$; QPC0 is located at $x = 0$ on edge S and at  $x = -d/2$ on edge $\alpha = 1$.
$\mathcal{H}_\text{island}$ is the Hamiltonian for the metallic island, as discussed in the main text.
The MZI is formed by the upper edge channel $\alpha = u$, lower edge channel $\alpha = d$, QPC1, and QPC2. The upper edge channel is described by  the Hamiltonian $\mathcal{H}_{\text{edge},u}$. The lower channel is the part of edge $\alpha =1$ outgoing from the metallic island, so its electron field $\psi_d (x)$ is identified as $\psi_d (x) = \psi_1(x+l+\frac{d}{2})$.
Electron tunneling at QPC1 and QPC2 is described by $\mathcal{H}_\text{MZI-QPC} = \mathcal{T}_1 + \mathcal{T}_2 + \text{h.c.}$, where the electron tunneling operators at the QPCs (from the lower to upper edge) are $\mathcal{T}_1 = \Gamma_1 \psi_{u}^\dagger(0)\psi_d(0)$ and $\mathcal{T}_2 = \Gamma_2\psi^\dagger_{u}(L_u)\psi_d(L_d)$ and h.c. means the hermitian conjugate; QPC1 is located at $x=0$ on both the upper and lower edges, while QPC2 is located at $x = L_u$ on the upper edge and at $x=L_d$ on the lower edge.
The tunneling amplitudes include the effect of the Aharonov-Bohm flux of an electron enclosed by the MZI loop, $\Gamma_1 = \gamma_1 e^{-i\pi \frac{\Phi}{\Phi_0}}$ and $\Gamma_2 = \gamma_2 e^{i\pi \frac{\Phi}{\Phi_0}}$.

To correctly describe electron tunneling events in the setup of the multiple edges, we attach a Klein factor~[S1] $F_\alpha$ to the electron operator of each edge channel $\alpha$ ($ = 1,\cdots, N$, S, $u$), $\psi^\dagger_{\alpha}(x) = F^\dagger_\alpha e^{-i\phi_\alpha(x)}/\sqrt{2\pi a}$,  where the Klein factors anticommute  $\{ {F_\alpha}, {F_\beta^\dagger} \} = 2\delta_{\alpha\beta}$, $\{ {F_\alpha}, {F_\beta} \} = 0$, and $[N_{\text{tot},\alpha},F_\beta] = -\delta_{\alpha\beta}F_\beta$. Here $N_{\text{tot},\alpha}$ is the operator counting the total number of electrons on the edge $\alpha$.

We calculate the interference current measured at the detector D in the tunneling regime (small $\gamma_i$'s) of the QPCs, employing the perturbation theory and Kelydsh Green's function. The current operator at the detector D is written as $I_\text{D} = e\dot{N_u}=ie(\langle \mathcal{T}_1 \rangle +\langle \mathcal{T}_2 \rangle  - \text{h.c.})$. It is decomposed, $I_\text{D} = I_\text{D}^{\text{dir}} +I_\text{D}^\text{int}$, into the direct current $I_\text{D}^{\text{dir}} \propto \gamma_0^2\gamma_1^2, \gamma_0^2\gamma_2^2$ and the interference current $I_\text{D}^\text{int} \propto \gamma_0^2\gamma_1\gamma_2$ that depends on $\Phi$.  
We focus on the counterclockwise winding of the interference current $\langle \mathcal{T}_1 \rangle^\text{int}$ and $\langle \mathcal{T}^\dagger_2 \rangle^\text{int}$ that is proportional to $e^{-i2\pi\Phi/\Phi_0}$; the winding of the opposite direction is obtained by the complex conjugation, $\langle \mathcal{T}^\dagger_1 \rangle^\text{int}$ and $\langle \mathcal{T}_2 \rangle^\text{int}$. We compute
\begin{equation}
\langle \mathcal{T}_1 (\tilde{t} =0) \rangle^\text{int}= -i\sum_{\eta ,\eta_1,\eta_2 = \pm 1} \eta\eta_1\eta_2 \int_{-\infty}^\infty dt_1^{\eta_1}dt_2^{\eta_2}dt^{\eta} \langle T_K\left[\mathcal{T}_1(\tilde{t}^+ = 0)\mathcal{T}_0(t_1^{\eta_1}) \mathcal{T}_0^\dagger (t_2^{\eta_2}) \mathcal{T}^\dagger_2(t^\eta)\right] \rangle.
\end{equation}
Here the time integral follows the contour of  $t_i \in (-\infty,\infty) \cup (\infty, -\infty)$, and we label $t_i = t_i^{\eta_i = \pm}$ if it is in the former (latter) branch of $(\mp \infty, \pm \infty)$. $T_K$ is the Keldysh contour ordering.
With some algebra, we get 
\begin{equation}\label{TVB-T1}
\begin{split}
\frac{\langle \mathcal{T}_1 (\tilde{t} =0)\rangle^\text{int}}{\gamma_0^2\gamma_1\gamma_2}= &\frac{-i}{(2\pi  v)^4}\sum_{\eta ,\eta_1,\eta_2 = \pm 1} \eta\eta_1\eta_2 \int _{-\infty}^0 dt \int_{-\infty}^\infty dt_1\int_{-\infty}^\infty dt_2\frac{e^{-i2\pi\Phi/\Phi_0} e^{-ieV (t_1-t_2)}}{\big[\frac{\beta}{\pi} \sin(\frac{\pi}{\beta} (\tau_0 + i\chi_{\eta_1, \eta_2} (t_1 - t_2)  (t_1 - t_2)))\big]^{2}}	\\
&\times  \frac{1}{\big[\frac{\beta}{\pi} \sin(\frac{\pi}{\beta} (\tau_0 - i\eta (t -L_d/v)))\big]\big[\frac{\beta}{\pi} \sin(\frac{\pi}{\beta} (\tau_0 - i\eta (t - L_u/v)))\big]} \\
&\times  \frac{\big[ \sin(\frac{\pi}{\beta} (\tau_0 - i\eta_1 (t_1-t +(d + L_d)/v)))\big]^{q} \big[ \sin(\frac{\pi}{\beta} (\tau_0 - i\eta_2 (t_2 +d/v)))\big]^{q}}{\big[ \sin(\frac{\pi}{\beta} (\tau_0 - i\eta_1 (t_1 +d/v)))\big]^{q} \big[ \sin(\frac{\pi}{\beta} (\tau_0 - i\eta_2 (t_2 -t +(d +L_d)/v)))\big]^{q}}\\
&\times  \frac{\big[ \sin(\frac{\pi}{\beta} (\tau_0 - i\eta_1 (t_1-t +(d + L_d+l)/v)))\big]^{1-q} \big[ \sin(\frac{\pi}{\beta} (\tau_0 - i\eta_2 (t_2 +(d+l)/v)))\big]^{1-q}}{\big[ \sin(\frac{\pi}{\beta} (\tau_0 - i\eta_1 (t_1 +(d+l)/v)))\big]^{1-q} \big[ \sin(\frac{\pi}{\beta} (\tau_0 - i\eta_2 (t_2 -t +(d +L_d+l)/v)))\big]^{1-q}},
\end{split}
\end{equation}
where $\chi_{\eta_1\eta_2}(t) = [(\eta_1+\eta_2)/2]\text{sgn}(t) - (\eta_1-\eta_2)/2$ and $\tau_0 = a/v$ is infinitesimal temporal cutoff.  The integrand in the first (second) line corresponds to the correlator of the electron that tunnels at QPC0 (QPC1 or QPC2). The third (fourth) line is attributed to the correlation of the electron tunneling at QPC1 or QPC2 and the charge (neutral) mode of the injected fractional excitation. As the neutral mode is not relevant (as it decays out before moving out of the coupling region), we put $l \rightarrow \infty$. Then, the last line from the neutral mode becomes 1. We note that the third line encodes all the information of the exchange between the electron $\psi^\dagger_1$ and the fractional excitation $\eta^\dagger_{1,q}$ in Eq.~(1) 
through the fractional expontent $q$.

Similarly, we compute  $-\langle \mathcal{T}_2^\dagger (t_0 = 0)\rangle^\text{int}$
and find that it has the same form with Eq. \eqref{TVB-T1} but with integration range $\int_0^\infty dt$. 
Collecting the results and their complex conjugates, we obtain the total interference current
\begin{equation}\label{inttodo}
\begin{split}
I_{D}^{\text{int}} & \propto  \gamma_0^2 \gamma_1 \gamma_2 \sum_{ \eta_1, \eta_2, \eta = \pm 1} \int_{-\infty}^{\infty} dt \int_{-\infty}^{\infty} dt_1 \int_{-\infty}^{\infty} dt_2 \eta_1 \eta_2 \eta e^{-i  2 \pi \Phi / \Phi_0} e^{-i eV (t_1 - t_2)} \mathcal{G} + h.c., \\
\mathcal{G} &= \frac{1}{\big[\frac{\beta}{\pi} \sin(\frac{\pi}{\beta} (\tau_0 + i\chi_{\eta_1, \eta_2} (t_1 - t_2)  (t_1 - t_2)))\big]^{2}} \frac{1}{\big[\frac{\beta}{\pi} \sin(\frac{\pi}{\beta} (\tau_0 - i\eta (t -L_d/v)))\big]\big[\frac{\beta}{\pi} \sin(\frac{\pi}{\beta} (\tau_0 - i\eta (t - L_u/v)))\big]}  \\
&\times \Bigg[ \frac{\big[ \sin(\frac{\pi}{\beta} (\tau_0 - i\eta_1 (t_1-t +(d + L_d)/v)))\big]^{q} \big[ \sin(\frac{\pi}{\beta} (\tau_0 - i\eta_2 (t_2 +d/v)))\big]^{q}}{\big[ \sin(\frac{\pi}{\beta} (\tau_0 - i\eta_1 (t_1 +d/v)))\big]^{q} \big[ \sin(\frac{\pi}{\beta} (\tau_0 - i\eta_2 (t_2 -t +(d +L_d)/v)))\big]^{q}}\Bigg].
\end{split}
\end{equation}
The contribution from the pole $t \sim L_u/v$ corresponds to the interference process $\langle s_2 |s_1 \rangle$, while  that from $t \sim L_d/v$ corresponds to $\langle s_2' |s_1 \rangle$. 
We evaluate the integral for large $eV>0$, where the integrand has a large peak at $t_1 \sim t_2 \equiv t_0 -d$. There, the second line is approximated as $\exp(-i\pi q\frac{\eta_1 -\eta_2}{2} \left[\text{sgn}\left(t_0 -t +L_d/v\right)-\text{sgn}\left(t_0\right)\right])$. For $\eta_1=\eta_2 = \pm 1$, it equals to 1, and $\langle s_2 |s_1 \rangle$ and $\langle s_2' |s_1 \rangle$ excactly cancel out each other.  The contribution from $\eta_1=-\eta_2 = -1$ corrresponds to the process originating from injection of a hole at QPC0, so is very small at low temperature. The major contribution comes from the process $\eta_1 = -\eta_2 = 1$  originating from injection of an electron at QPC0,
\begin{equation}
\begin{split}
\frac{I_{D}^{\text{int}}}{\gamma_0^2 \gamma_1 \gamma_2} & \propto -2\pi eV  \sum_{ \eta = \pm 1}\eta  \int_{-\infty}^\infty dt_0\int_{-\infty}^\infty dt 		\frac{e^{-i  2 \pi \Phi / \Phi_0}e^{-i\pi q\left[\text{sgn}\left(t_0 -t +L_d/v\right)-\text{sgn}\left(t_0\right)\right]}}{\big[\frac{\beta}{\pi} \sin(\frac{\pi}{\beta} (\tau_0 - i\eta (t -L_d/v)))\big]\big[\frac{\beta}{\pi} \sin(\frac{\pi}{\beta} (\tau_0 - i\eta (t - L_u/v)))\big]}	\\
& \propto i\frac{(2\pi)^2 eV e^{-i  2 \pi \Phi / \Phi_0}}{\sinh(\pi (L_d-L_u)/\beta v)}  \int_{-\infty}^\infty dt_0	(e^{-i\pi q\left[\text{sgn}\left(t_0 -(L_u-L_d)/v\right)-\text{sgn}\left(t_0\right)\right]}-1) +h.c.,	
\end{split}
\end{equation}
where the first term of the integrand in the second line corresponds to the process $\langle s_2' |s_1 \rangle$ from the pole at $t \sim L_d/v$, and the second term corresponds to the process $\langle s_2 |s_1 \rangle$ from the pole at $t \sim L_u/v$. For $L_d-L_u >0$, the latter becomes a non-trivial phase factor of $e^{- i2\pi q} $ for $-\Delta L/v<t_0 <0 $. In the opposite case of $L_d-L_u <0$, it becomes $e^{ i2\pi q} $ for $0<t_0 <\Delta L/v $. As a result, we get
\begin{equation}
I_D^\text{int} =  \gamma_0^2 \gamma_1 \gamma_2	\frac{\Delta L}{v}\frac{eV }{\pi \beta v^4\sinh(\pi \Delta L/\beta v)} \sin \pi q\cos(2\pi \frac{\Phi}{\Phi_0} + \pi q\text{sgn}(L_d-L_u))
\label{DEfinal}
\end{equation}
reproducing Eq.~(5). 

\begin{figure}[t!]
	\centering
	\includegraphics[width = 0.7 \textwidth]{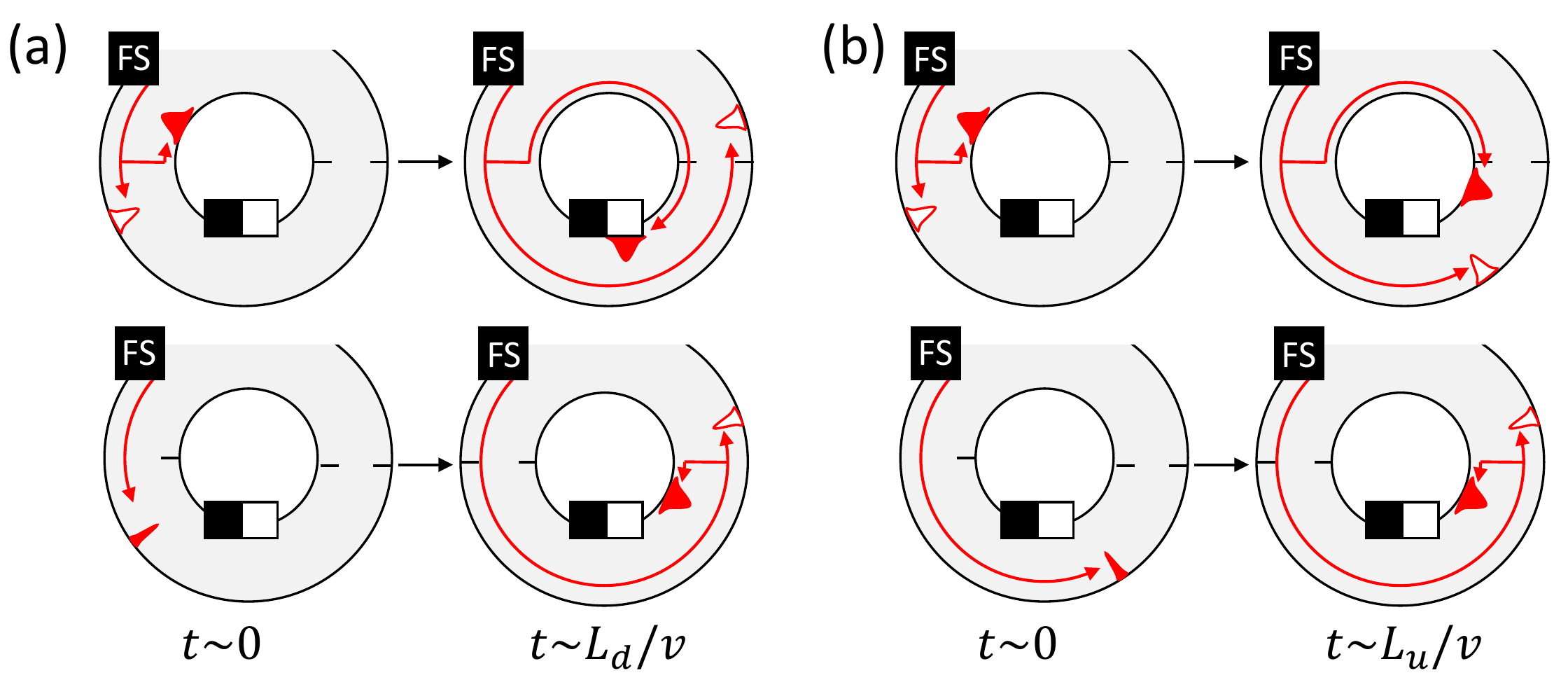}
	\caption{Two conventional interference processes (a) and (b) in which electron tunneling (thick red peaks) happens at QPC1 or QPC2, leaving a fractional hole (empty thinner peaks) behind, at the time when the injected fractional charge arrives at the QPC. 
		The contribution of these conventional processes to the interference current becomes negligible when the arm-length difference $\Delta L$ is larger than the spatial width $w$ of the fractional charge (equivalently when the voltage $V$ is sufficiently large).}\label{figS-conv}
\end{figure}

In the above calculation at large $eV$, the interference involving the double exchange between the fractional charge and an electron dominates over the conventional interference in which electron tunneling happens at QPC1 or QPC2, leaving a fractional hole behind, at the time when the injected fractional charge arrives at the QPC; the fractional charge is directly involved in the conventional interference.
We estimate the contribution of the conventional process to the interference current in Eq.~\eqref{inttodo}  at large $V$.
choosing the branch cuts of $t_1 \sim -d/v$ and $t_2 \sim t - (d+L_d)/v$ in \eqref{inttodo} and using $\int_{-\infty}^\infty dt e^{-ieVt } \left( \frac{\beta}{\pi} \sin (\frac{\pi}{\beta}(\tau_0 -it))\right)^{-\nu} \simeq \frac{2\pi }{\Gamma(\nu)}(eV)^{\nu-1}$, we compute the contribution of the conventional process from Eq.~\eqref{inttodo}, 
\begin{equation}
\begin{split}
-\sum_\eta \eta\int_{-\infty}^\infty dt \frac{(2\pi)^2eV^{2q - 2}e^{-i  2 \pi \Phi / \Phi_0} / \Gamma(q)^2}{\big[\frac{\beta}{\pi} \sin(\frac{\pi}{\beta} (\tau_0 - i\eta (t -L_d/v)))\big]\big[\frac{\beta}{\pi} \sin(\frac{\pi}{\beta} (\tau_0 - i\eta (t - L_u/v)))\big]}\frac{e^{ieV (t-L_d/v)}}{\big[\frac{\beta}{\pi} \sin(\frac{\pi}{\beta} (\tau_0 + i (t - L_d/v)))\big]^{2-2q}}.
\end{split}
\end{equation}
This is further evaluated by using the contour integral. The processes around $t \sim L_u/v +i\tau_0$ and $t \sim L_d/v +i\tau_0$ have the contribution of
\begin{eqnarray}
I_{D,\text{conv (a)}}^{\text{int}}  & \propto & \gamma_0^2 \gamma_1 \gamma_2 \frac{2(2\pi)^3}{\Gamma(q)^2\Gamma(3-2q)} 
\frac{eV}{\frac{\beta eV}{\pi}\sinh(\pi \Delta L/\beta v)} \cos(2\pi\frac{\Phi}{\Phi_0} + \frac{\pi}{2}\text{sgn}(L_d-L_u)), \\
I_{D,\text{conv (b)}}^{\text{int}} &\propto & \gamma_0^2 \gamma_1 \gamma_2\frac{2(2\pi)^3}{\Gamma(q)^2 } 	 \frac{eV}{[\frac{\beta eV}{\pi}\sinh(\pi \Delta L/\beta v)]^{3-2q}}\cos(2\pi \frac{\Phi}{\Phi_0} - \frac{3-2q}{2}\pi\text{sgn}(L_d-L_u) + \frac{eV(L_d-L_u)}{v} ),
\label{CONV-b}
\end{eqnarray}
respectively. Each process is illustrated in Fig. \ref{figS-conv} (a) and (b). 
At large $V$, both of $I_{D,\text{conv (a)}}^{\text{int}}$ and $I_{D,\text{conv (b)}}^{\text{int}}$ are smaller than the result in Eq.~\eqref{DEfinal} coming from our main interference process with the double exchange between a fractional charge and an electron.
It is more convenient to see the differential conductance $dI^\text{int}/dV$ (rather than $I^\text{int}$), to detect the phase shift $\pi q$ in  Eq.~\eqref{DEfinal}; the phase shift converges more quickly to the desired phase shift $\pi q$ as $V$ increases. This can be seen from the ratio of the differential conductance between the conventional-process contribution and our main process contribution,
\begin{equation}\label{compare}
\frac{\text{Conventional Process }}{\text{Double Exchange Process}}=  \mathcal{O}\left((\frac{w}{L_\beta})^{2-2q}\frac{w}{\Delta L}\csch(\frac{\Delta L}{L_\beta})^{2-2q}\right),
\end{equation}
which is negligible for $w \ll \Delta L, L_\beta$.

Figure~\ref{TVBsetup}(b) shows numerical calculation of the phase shift resulting from all the processes discussed above.
The phase shift approaches the fractional exchange phase $\pm \pi q$ at large $\Delta L$ and large bias voltage $V$, accompanied by a small oscillation.
The small oscillation is caused by the dynamical phase which a fractional excitation gains in the conventional interference process [see the last phase term in Eq.~\eqref{CONV-b}], and its period is inversely proportional to the bias voltage.
The small oscillation is suppressed at large $\Delta L / w$, where the main interference process with the double exchange dominates over the conventional process [see Eq.~\eqref{compare}]. 
Note that the net dynamical phase of a fractional excitation is zero in the main interference process, not affecting the phase shift; the dynamical phase gained by a fractional excitation in one subprocess of the main interference cancels that of the other subprocess, since the fractional excitation propagates the same distance in the two subprocesses.

We note that when an electron (instead of a fractional charge) is injected to the MZI, our main interference process with the double exchange vanishes and the conventional process becomes dominant. In this case, the phase shift linearly increases with $\Delta L$ due to the dynamical phase.

\section{MZI combined with voltage pulse}

We provide detailed derivation of  Eq.~(6) of the main text.
It is the time-averaged interference current of a MZI combined with the fractional-charge source formed by voltage pulse. 
The Hamiltonian is written as 
\begin{equation}
\begin{split}
\mathcal{H}= & \mathcal{H}_V(t) + \sum_{\alpha = u,d}\mathcal{H}_{\text{edge},\alpha}	+ \mathcal{H}_\text{MZI-QPC}.		\\
\end{split}
\end{equation}
Here, $\mathcal{H}_V(t)=\frac{e}{2\pi}V(t)\int_{-\infty}^0 \partial_x\phi_d(x,t)$ describes the coupling between the voltage pulse and the electron density of the lower edge channel of the MZI (the voltage pulse is assumed to be applied at $x < 0$). And the last two terms of the Hamiltonian is for the MZI ($\mathcal{H}_\text{MZI-QPC}$ was introduced in the previous section). For convenience, we consider a Lorentzian voltage pulse, $V(t) = \frac{V_0}{\pi}\sum_k \frac{w/(vT)}{(w/(vT))^2 + (t/T-k)^2}$, where $T$ and $w$ are the period and width of the pulse respectively, and $e^2V_0/h = q$ defines the amount of the fractional charge generated per period. 
Thanks to the chirality and linearity of the quantum Hall edge channel, the generated fractional charge with Loretnzian spatial distribution will propagate without distortion although it is composed of many particle-hole pairs [S2]. The solution of the equation of motion of the boson field $\phi_d$ of the lower edge is $\phi_d(x,t) = \phi_d^{(0)}(x,t) + \varphi(x,t)$, where $\phi_d^{(0)}(x,t)$ is the field without any voltage bias, and $\varphi(x,t) = e\int_{-\infty}^t dt' V(x/v-(t-t'))$ is the phase shift due to the bias,
\begin{equation}
e^{i\varphi(x,t)}=  \prod_k \frac{[t-x/v +i(w/v +iTk)]^q}{[t-x/v -i(w/v -iTk)]^q}.
\end{equation}
$e^{i\varphi(x,t)}$ carries the information of the statistics of the fractional charge.


\begin{figure}[t!]
	\centering
	\includegraphics[width = 1.0 \textwidth]{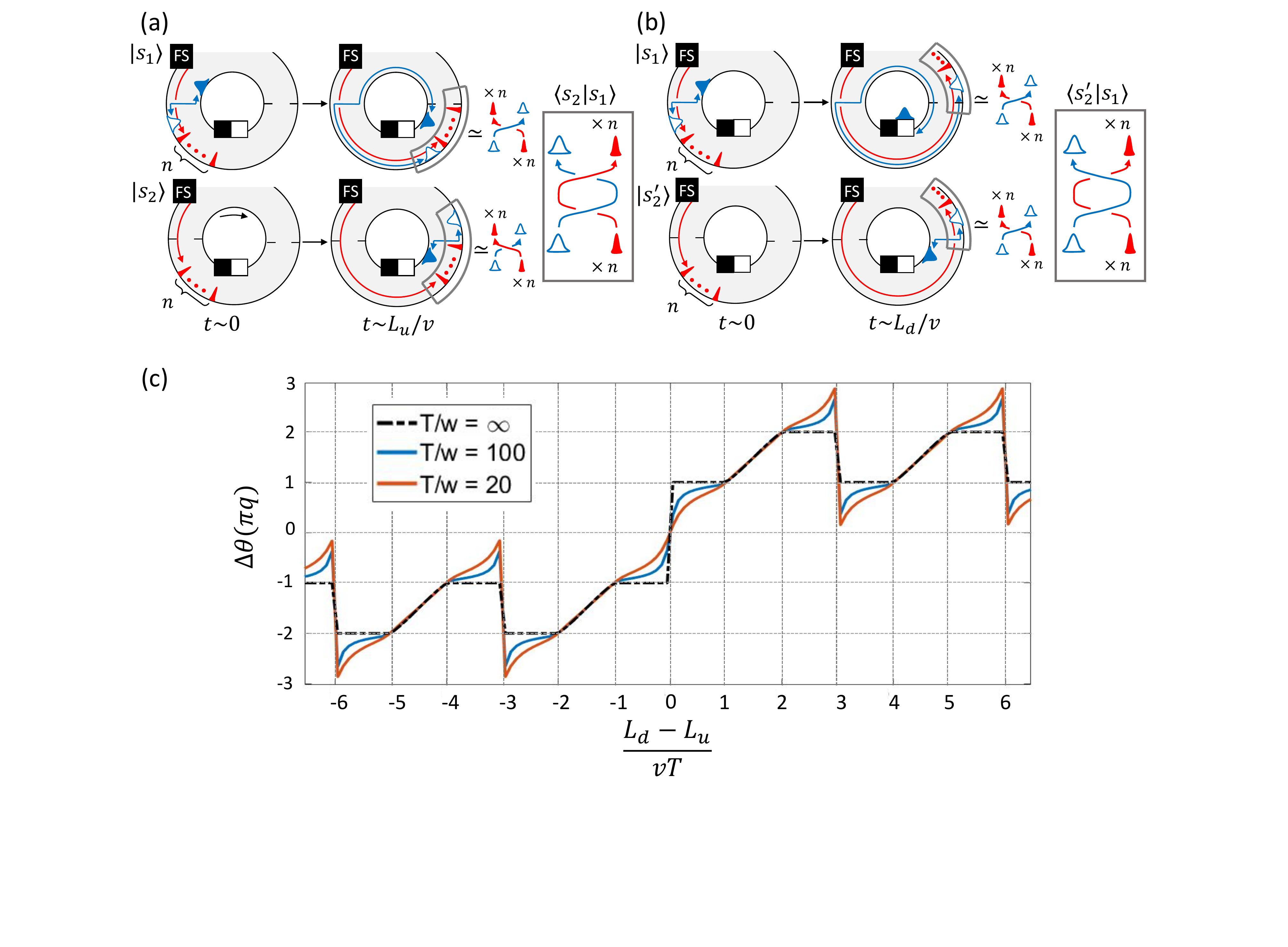}
	\caption{ (a) Interference process $\langle s_2 | s_1 \rangle$  in which an electron has double exchange with multiple fractional charges. (b)
		Interference process $\langle s'_2 | s_1 \rangle$ where the double exchange does not occur. (c) Interference phase shift due to the two processes. The result is obtained with temperature 20 mK, $T=100$ ps,  $q = 1/3$, and $v = 10^5$ m/s.}\label{figS-voltt}
\end{figure}

In the tunneling regime of the two QPCs of the MZI, the time-averaged current at detector D can be calculated using linear response theory $\overline{I_D} \simeq i\int_0^T \frac{d\tilde{t}}{T} \int_{-\infty}^{\tilde{t}} dt \langle [{I_D(\tilde{t})},\mathcal{H}_\text{MZI}(t)]\rangle$. 
Its interference part $\overline{I_D^\text{int}} \propto \gamma_1\gamma_2$ that depends on the Aharonov-Bohm flux $\Phi$ enclosed by the MZI loop is decomposed into $\overline{I_D^\text{int}} \simeq i(\overline{\mathcal{T}_1^\text{int}} +\overline{\mathcal{T}_2^{ \text{int}} } -h.c.)$. 
The first term $\overline{\mathcal{T}_i^\text{int}} = i\int_0^T  \frac{d\tilde{t}}{T} \int_{-\infty}^{\tilde{t}} dt \langle [\mathcal{T}_i(\tilde{t}), \mathcal{T}_{j \neq i}^\dagger(t)]\rangle$ is computed as
\begin{equation}
\begin{split}
&\overline{\mathcal{T}_1^\text{int}}=  \int_{0}^T \frac{d\tilde{t}}{T} \langle \mathcal{T}_1(\tilde{t})\rangle^\text{int} = i\int \frac{d\tilde{t}}{T} \int_{-\infty}^{\tilde{t}} dt \langle [\mathcal{T}_1(\tilde{t}), \mathcal{T}_{2}^\dagger(t)]\rangle		\\
= & i\gamma_1\gamma_2\sum_{\eta = \pm}\eta\int_{0}^T \frac{d\tilde{t}}{T} \int_{-\infty}^{\tilde{t}} dt \langle [ \psi_{d}^{(0)}(0,\tilde{t})\psi_{u}^{(0)\dagger}(0,\tilde{t}), \psi_{d}^{(0)\dagger}(L_d, t)\psi_{u}^{(0)}(L_u, t)] \rangle e^{-i  2 \pi\frac{ \Phi }{ \Phi_0}} e^{i\varphi(0,\tilde{t})} e^{-i\varphi(L_d,t)} \\
= & \frac{i\gamma_1\gamma_2}{(2\pi v)^2}\sum_{\eta = \pm}\eta\int_{0}^T \frac{d\tilde{t}}{T} \int_{-\infty}^{\tilde{t}} dt \frac{e^{-i  2 \pi \Phi / \Phi_0}e^{i\varphi(0,\tilde{t})} e^{-i\varphi(L_d,t)}  }{\big[\frac{\beta}{\pi} \sin(\frac{\pi}{\beta} (\tau_0 - i\eta (t-\tilde{t} -L_d/v)))\big]\big[\frac{\beta}{\pi} \sin(\frac{\pi}{\beta} (\tau_0 - i\eta (t-\tilde{t} - L_u/v)))\big]}. \\
\end{split}
\end{equation}
Computing the other terms in the same way, we get the interference current,
\begin{equation}\label{intmain}
\begin{split}
\overline{I_D^\text{int}} \propto -\gamma_1\gamma_2\sum_{\eta= \pm} \eta\int_{0}^T \frac{d\tilde{t}}{T} \int_{-\infty}^{\infty}dt \frac{e^{-i  2 \pi \Phi / \Phi_0}  e^{i\varphi(0,\tilde{t})} e^{-i\varphi(L_d,t)} }{\big[\frac{\beta}{\pi} \sin(\frac{\pi}{\beta} (\tau_0 - i\eta (t-\tilde{t} -L_d/v)))\big]\big[\frac{\beta}{\pi} \sin(\frac{\pi}{\beta} (\tau_0 - i\eta (t-\tilde{t} - L_u/v)))\big]} + h.c.
\end{split}
\end{equation}
In the limit of $w \ll vT$, $e^{i\varphi(0,\tilde{t})} e^{-i\varphi(L_d,t)}$  becomes the form of $e^{-i2n\pi q}, n \in \mathbb{Z}$ when $- n T< t-L_d/v < (1-n)T$. In this time domain, an electron has double exchange with $n$ fractional charges in its interference process.
Then, the interference current becomes
\begin{equation}
\begin{split}
\frac{\overline{I_D^\text{int}}}{\gamma_1\gamma_2} \propto -\sum_{\eta= \pm, n \in \mathbb{Z}} \eta\int_{0}^T \frac{d\tilde{t}}{T} \int_{-nT+L_d/v}^{(1-n)T+L_d/v}dt \frac{e^{-i  2 \pi \Phi / \Phi_0} e^{-i2n\pi q}}{\big[\frac{\beta}{\pi} \sin(\frac{\pi}{\beta} (\tau_0 - i\eta (t-\tilde{t} -L_d/v)))\big]\big[\frac{\beta}{\pi} \sin(\frac{\pi}{\beta} (\tau_0 - i\eta (t-\tilde{t} - L_u/v)))\big]} + h.c.
\end{split}
\end{equation}
The process $\langle s_2' |s_1 \rangle$ comes from the pole at $t \sim \tilde{t} +L_d/v$ and always corresponds to $n=0$, namely, it does not include any double exchange. On the other hand, the process $\langle s_2 |s_1 \rangle$ comes from the pole at $t \sim \tilde{t} +L_u/v$ and has double exchange (braiding) with fractional charges, the number of which depends on the value of $\tilde{t}$. The number of the braided fractional charges is $n = \lfloor \frac{L_d-L_u}{vT} \rfloor$ when  $\frac{L_d-L_u}{v}-\lfloor \frac{L_d-L_u}{vT} \rfloor T<\tilde{t}< T   $, and $n =\lfloor \frac{L_d-L_u}{vT} \rfloor+1$ when  $ 0<\tilde{t}<  \frac{L_d-L_u}{v}-\lfloor \frac{L_d-L_u}{vT} \rfloor T $ . The integration of $t$ and $\tilde{t}$ gives

\begin{equation}
\begin{split}
\overline{I_D^\text{int}} =  \frac{\pm  i\gamma_1\gamma_2e^{-i  2 \pi \Phi / \Phi_0}}{2v^2\beta T\sinh(\pi \Delta L/\beta v)}\left( e^{ \mp i2\pi q \lfloor \frac{\Delta L}{vT} \rfloor}\left(\lceil \frac{\Delta L}{vT} \rceil T-\frac{\Delta L}{v} \right) + e^{\mp i2\pi q (\lfloor \frac{\Delta L}{vT} \rfloor +1)}\left(T +\frac{\Delta L}{v}-\lceil \frac{\Delta L}{vT} \rceil T \right)-T \right) + h.c.
\end{split}
\end{equation}

Rearranging this, we obtain Eq.~(6) of the main text. 


In Fig.~\ref{figS-voltt}, the phase shift of the time-averaged interference current is drawn for the case $q=1/3$.
For small arm-length difference $\Delta L \leq vT$, the phase shift saturates to $\pm \pi q$, as shown in Fig.~2(b) of the main text.
For  $1 < \Delta L/(vT) \leq 2$, the phase shift linearly changes from $\pm \pi q$ to $\pm 2\pi q$, and is attributed to a mixture of the cases with $n=1$ and $n=2$ braided fractional charges. 
For  $2 < \Delta L/(vT) \leq 3$, the number of braided fractional charges is $n=2$ or $n=3$. 
In this case, the phase shift saturates to $\pm 2 \pi q$, since $\langle s_2 |s_1 \rangle$ fully cancels with $\langle s_2' |s_1 \rangle$ (or $e^{-i6\pi q} = 1$) when $n=3$ (three fractional charges are braided by an electron). 
If $\Delta L / (vT)$ further increases, the phase shift changes periodically, with alternating plateaus of $\pm \pi q$ and $\pm 2\pi q$. The plateaus are clearly visible for large $T/w$, and are smoothed for small $T/w$.

\end{document}